\documentclass{llncs}
\usepackage{makeidx}  

\usepackage{amssymb}
\usepackage{graphicx}
\usepackage{amsmath}
\usepackage{url}

\usepackage{enumitem}
\usepackage{mymacros}
\usepackage{algorithm}
\usepackage{algpseudocode}
\usepackage{booktabs}  
\usepackage{siunitx}

\usepackage{multirow}
\usepackage{wrapfig}

 \def\sact{\triangleright}
 \def\eact{{\scriptstyle\square}}
 \newcommand*\rfrac[2]{{}^{#1}\!/_{#2}}

\begin{document}
\mainmatter              
\title{Learning Hybrid Process Models From Events}
\subtitle{Process Discovery Without Faking Confidence \\
(Experimental Results)\footnote{This technical report complements the paper ``W.M.P. van der Aalst, R. De Masellis, C. Di Francescomarino, C. Ghidini, \emph{Learning Hybrid Process Models From Events}'' submitted to the BPM 2017 conference.}}
\titlerunning{Learning Hybrid Process Models From Events}  
%
\author{Wil M.P. van der Aalst\inst{1,2} and Riccardo De Masellis\inst{2} and Chiara Di Francescomarino\inst{2} and Chiara Ghidini\inst{2}}
\authorrunning{Wil M.P. van der Aalst}   
%
%
\institute{Eindhoven University of Technology, PO Box 513, Eindhoven, The Netherlands\and
FBK-IRST, Via Sommarive 18, 38050 Trento, Italy}

\maketitle              

\begin{abstract}        
Process discovery techniques return process models that are either formal
(precisely describing the possible behaviors)
or informal (merely a ``picture'' not allowing for any form of formal reasoning).
Formal models are able to classify traces
(i.e., sequences of events) as fitting or non-fitting.
Most process mining approaches described in the literature produce such models.
This is in stark contrast with the over 25 available
commercial process mining tools that only discover informal process models
that remain \emph{deliberately vague} on the precise set of possible traces.
There are two main reasons why vendors resort to such models:
\emph{scalability} and \emph{simplicity}.
In this paper, we propose to combine the best of both worlds:
discovering \emph{hybrid process models} that have formal and informal elements.
As a proof of concept we present a discovery technique based on \emph{hybrid Petri nets}.
These models allow for formal reasoning, but also reveal information that cannot be
captured in mainstream formal models.
A novel discovery algorithm returning hybrid Petri nets has been implemented in ProM and
has been applied to several real-life event logs.
The results clearly demonstrate the advantages of remaining ``vague''
when there is not enough ``evidence'' in the data or standard modeling constructs do not ``fit''.
Moreover, the approach is scalable enough to be incorporated in industrial-strength process mining tools.
\end{abstract}

\section{Introduction}
\label{sec:intro}

The increased interest in process mining illustrates that Business Process Management (BPM) is rapidly becoming more data-driven \cite{process-mining-book-2016}.
Evidence-based BPM based on process mining helps to create a common ground for business process improvement and information systems development.
The uptake of process mining is reflected by the growing number of commercial process mining tools available today.
There are over 25 commercial products supporting process mining (Celonis, Disco, Minit, myInvenio, ProcessGold, QPR, etc.).
All support process discovery and can be used to improve compliance and performance problems. For example,
without any modeling, it is possible to learn process models clearly showing the main bottlenecks and deviating behaviors.

These commercial tools are based on variants of
techniques like the heuristic miner \cite{aal_min_icae} and the fuzzy miner \cite{fuzzy_BPM2007}
developed over a decade ago \cite{process-mining-book-2016}.
All return process models that \emph{lack formal semantics} and thus cannot be used as a \emph{classifier} for traces.
Classifying traces into \emph{fitting} (behavior allowed by the model)
and \emph{non-fitting} (not possible according to the model) is however important for more advanced types of process mining.
Informal models (``boxes and arcs'') provide valuable insights, but cannot be used to draw reliable conclusions.
Therefore, most discovery algorithms described in the literature
(e.g., the  $\alpha$-algorithm \cite{aal_min_TKDE}, the region-based approaches \cite{lorenz_BPM2007,carmona-PN2010,language_mining_dongen-fundamenta2009}, and the inductive mining approaches \cite{sander-tree-disc-PN2013,sander-infreq-bpi2013-lnbip2014,sander-scalable-procmin-SOSYM})
produce formal models (Petri nets, transition systems, automata, process trees, etc.) having clear semantics.

So why did vendors of commercial process mining tools opt for informal models?  Some of the main drivers for this choice include:
\begin{itemize}
  \item \emph{Simplicity}: Formal models may be hard to understand. End-users need to be able to interpret process mining results: Petri nets with smartly constructed places and BPMN with many gateways are quickly perceived as too complex.
  \item \emph{Vagueness}: Formal models act as binary classifiers: traces are fitting or non-fitting. For real-life processes this is often not so clear cut.
  The model capturing 80 percent of all traces may be simple and more valuable than the model that allows for all outliers and deviations seen in the event log.
  Hence, ``vagueness'' may be desirable to show relationships that cannot be interpreted in a precise manner.
  \item \emph{Scalability}: Commercial process mining tools need to be able to handle logs with millions of events and still be used in an interactive manner.
  Many of the more sophisticated discovery algorithms producing formal models (e.g., region-based approaches  \cite{lorenz_BPM2007,carmona-PN2010,language_mining_dongen-fundamenta2009}) do not scale well.
\end{itemize}
The state-of-the-art commercial products show that simplicity, vagueness and scalability can be combined effectively.
Obviously, vagueness and simplicity may also pose problems.
People may not trust process mining results when a precise interpretation of the generated model is impossible.
When an activity has multiple outgoing arcs, i.e., multiple preceding activities, one would like to know whether these are concurrent or in a choice relation.
Which combinations of output arcs can be combined? Showing frequencies on nodes (activities) and arcs may further add to the confusion when ``numbers do not add up''.

We propose \emph{hybrid process models} as a way to combine the best of both worlds.
Such models show informal dependencies (like in commercial tools) that are deliberately vague and at the same time provide formal semantics for the parts that are clear-cut. Whenever there is enough structure and evidence in the data, explicit routing constructs are used.
If dependencies are weak or too complex, then they are not left out, but depicted in an informal manner.

We use \emph{hybrid Petri nets}, a new class for Petri nets with informal annotations, as a concrete representation of hybrid process models.
However, the ideas, concepts, and algorithms are generic and could also be used in the context of BPMN, UML activity diagrams, etc.
Our proposed \emph{discovery technique} has two phases. First we discover a \emph{causal graph} based on the event log.
Based on different (threshold) parameters we scan the event log for possible causalities.
In the second phase we try to learn places based on explicit quality criteria.
Places added can be interpreted in a precise manner and have a guaranteed quality.
Causal relations that cannot or should not be expressed in terms of places are added as sure or unsure arcs.
The resulting hybrid Petri net can be used as a starting point for other types of process mining.

The approach has been implemented in \emph{ProM} and has been tested on various event logs and processes.
These applications of our approach show that hybrid process models are useful and combine the best of both worlds:
\emph{simplicity, vagueness, and scalability can be combined with partly formal models that allow for reasoning and provide formal guarantees}.

The remainder is organized as follows.
We first present a running example (Sect.~\ref{sub:motexample}) and some preliminaries (Sect.~\ref{subsec:prelim}).
Sect.~\ref{sec:hybridPetrinet} defines hybrid Petri nets.
The actual two-phase discovery approach is presented in Sect.~\ref{sec:disc}.
Sect.~\ref{sec:impl} describes the \emph{ProM} plug-ins developed to support the discovery of hybrid process models.
Sect.~\ref{sec:eval} evaluates the approach.
Sect.~\ref{sec:relwork} discusses related work and Sect.~\ref{sec:concl} concludes the paper.

\section{Motivating Example}
\label{sub:motexample}

\begin{figure}[t!]
\centerline{\includegraphics[width=12cm]{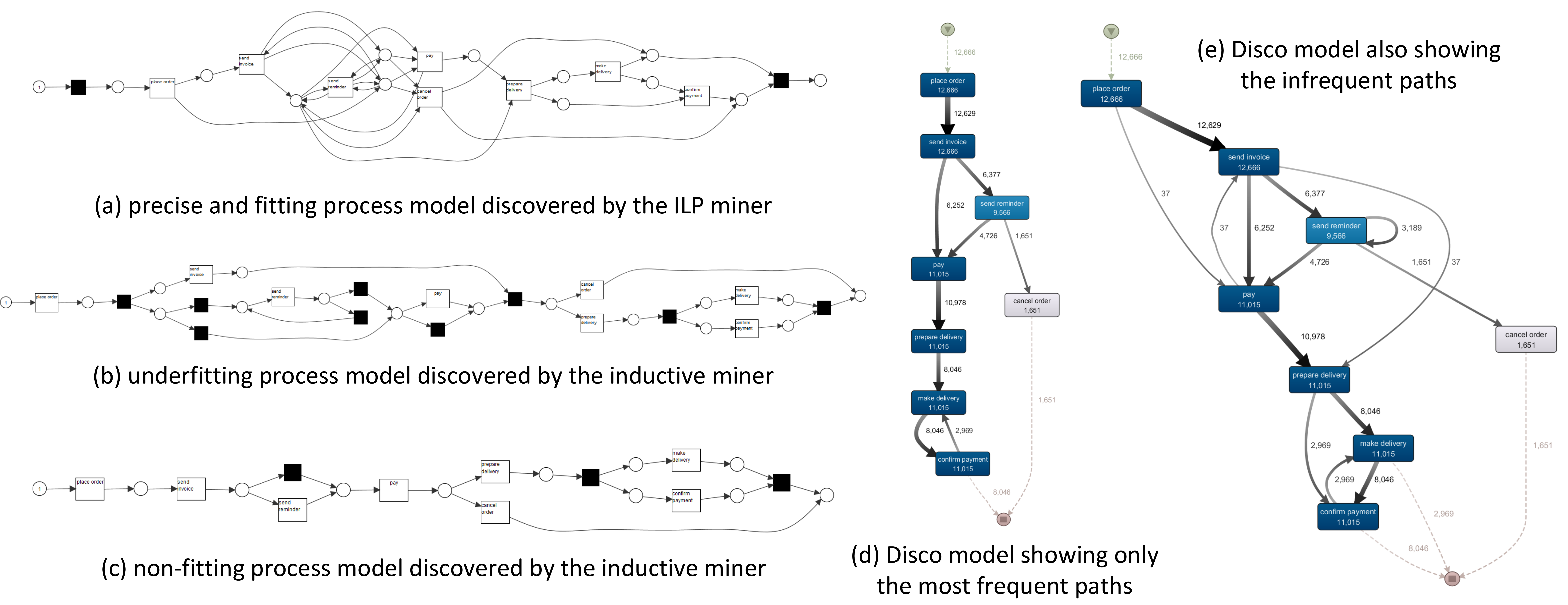}}
\caption{Five process models discovered for an event log recorded for 12,666 orders (labels are not intended to be readable).}
\label{f-intro-prom-disco}
\end{figure}

Figure~\ref{f-intro-prom-disco} illustrates the trade-offs using example data from an order handling process.
All five models have been produced for the same event log containing 12,666 cases, 80,609 events, and eight unique activities.
Each case has a corresponding trace, i.e., a sequence of events.
Models (a), (b), and (c) are expressed in terms of a Petri net and have formal semantics.
Model (a) was created using the ILP miner with default settings; it is precise and each of the 12,666 cases perfectly fits the model.
However, model (a) is difficult to read. For larger event logs, having more activities and
low-frequent paths, the ILP miner is not able to produce meaningful models (the approach becomes intractable and/or produces incomprehensible models).
Models (b) and (c) were created using the inductive miner (IMf \cite{sander-infreq-bpi2013-lnbip2014}) with different settings for the noise threshold (0.0 respectively 0.2). Model (b) is underfitting, but able to replay all cases.
Model (c) focuses on the mainstream behavior only, but only 9,440 of the 12,666 cases fit perfectly.
In 3,189 cases there are multiple reminders and in 37 cases the payment is done before sending the invoice. All other cases conform to model (c).
Models (d) and (e) were created using the commercial process mining tool \emph{Disco} (Fluxicon) using different settings.
These models are informal. Model (d) shows only the most frequent paths and model (e) shows all possible paths.
For such informal models it is impossible to determine the exact nature of splits and joins.
Commercial tools have problems dealing with loops and concurrency. For example, for each of the 12,666 cases, activities \emph{make delivery} and \emph{confirm payment}
happened at most once, but not in a fixed order. However, these concurrent activities are put into a loop in models (d) and (e).
This problem is not specific for \emph{Disco} or this event log: all commercial tools suffer from this problem.

We would like to combine the left-hand side and the right-hand side of Figure~\ref{f-intro-prom-disco}
by using formal semantics when the behavior is clear and easy to express and
resorting to informal annotations when things are blurry or inexact.

\section{Preliminaries}
\label{subsec:prelim}

In this section we introduce basic concepts, including multisets, operations on sequences, event logs and Petri nets.

$\bag(A)$ is the set of all multisets over some set $A$.
For some multiset $X\in \bag(A)$, $X(a)$ denotes the number of times element $a\in A$ appears in $X$.
Some examples: $X = [\,]$, $Y = [x,x,y]$, and $Z = [x^3,y^2,z]$
are multisets over $A=\{x,y,z\}$.
$X$ is the empty multiset, $Y$ has three elements ($Y(x) = 2$, $Y(y) = 1$, and $Y(z) = 0$), and $Z$ has six elements.
Note that the ordering of elements is irrelevant.

$\sigma = \langle a_1,a_2, \ldots, a_n\rangle \in A^*$ denotes a sequence over $A$.
$\sigma(i) = a_i$ denotes the $i$-th element of the sequence.
$\card{\sigma} = n$ is the length of $\sigma$ and $\mi{dom}(\sigma) = \{1, \ldots, \card{\sigma}\}$ is the domain of $\sigma$.
$\langle\,\rangle$ is the empty sequence, i.e., $\card{ \langle\,\rangle } = 0$ and  $\mi{dom}(\langle\,\rangle) = \emptyset$.
$\sigma_1 \cdot \sigma_2$ is the concatenation of two sequences.

Let $A$ be a set and $X\subseteq A$ one of its subsets.
$~\tproj_{X} \in A^* \rightarrow X^*$ is a projection function and is defined recursively:
$\langle\,\rangle \tproj_{X} = \langle\,\rangle$ and
for $\sigma \in A^*$ and $a\in A$:
$(\langle a \rangle \cdot \sigma)\tproj_{X} = \sigma \tproj_{X}$ if $a \not\in X$ and
$(\langle a \rangle \cdot \sigma)\tproj_{X} = \langle a \rangle \cdot \sigma \tproj_{X}$ if $a \in X$.
For example, $\langle a,b,a\rangle \tproj_{\{a,c\}} = \langle a,a\rangle$.
Projection can also be applied to multisets of sequences, e.g.,
$[\langle a,b,a\rangle^5,\langle a,d,a\rangle^5,\langle a,c,e\rangle^3] \tproj_{\{a,c\}} =
[\langle a,a\rangle^{10},\langle a,c\rangle^3]$.

Starting point for process discovery is an event log where events are grouped into cases.
Each case is represented by a trace, e.g., $\langle \sact,a,b,c,d,\eact \rangle$.

\begin{definition}[Event Log]\label{def:eventlog}
An \emph{event log} $L \in \bag(A^*)$ is a non-empty multiset of traces over some activity set $A$.
A trace $\sigma \in L$  is a sequence of activities. There is a special start activity $\sact$ and a special end activity $\eact$.
We require that $\{\sact,\eact\} \subseteq A$ and each trace $\sigma \in L$ has the structure
$\sigma = \langle \sact, a_1, a_2, \ldots, a_n, \eact \rangle$ and $\{\sact,\eact\} \cap \{ a_1, a_2, \ldots, a_n\} = \emptyset$.
${\cal U}_{L}$ is the set of all event logs satisfying these requirements.
\end{definition}

An event log captures the observed behavior that is used to learn a process model.
An example log is $L_1 = [
\langle \sact,a,b,c,d,\eact \rangle^{45},
\langle \sact,a,c,b,d,\eact \rangle^{35},
\langle \sact,a,e,d, \allowbreak \eact \rangle^{20}]$
containing 100 traces and 580 events.
In reality, each event has a timestamp and may have any number of additional attributes.
For example, an event may refer to a customer, a product, the person executing the event, associated costs, etc.
Here we abstract from these notions and simply represent an event by its activity name.

A \emph{Petri net} is a bipartite graph composed of places (represented by circles) and transitions (represented by squares).
\begin{definition}[Petri Net]
A Petri net is a tuple $N=(P,T,F)$ with $P$ the set of places, $T$ the set of transitions,
$P \cap T = \emptyset$, and $F\subseteq (P \times T) \cup (T \times P)$ the flow relation.
\end{definition}

Transitions represent activities and places are added to model causal relations.
$\pre x = \{y \mid (y,x) \in F\}$ and $\post x = \{y \mid (x,y) \in F\}$ define input and output sets of places and transitions.
Places can be used to causally connect transitions as is reflected by relation $\widehat{F}$: $(t_1,t_2) \in \widehat{F}$ if $t_1$ and $t_2$ are connected through a place $p$,
i.e., $p \in \post{t_1}$ and $p \in \pre{t_2}$.

\begin{definition}[$\widehat{F}$]\label{def:hat}
Let $N=(P,T,F)$ be a Petri net.
$\widehat{F} = \{(t_1,t_2)\in T \times T \mid \exists_{p \in P}\ \{(t_1,p),(p,t_2)\} \subseteq F\}$ are all pairs of transitions connected through places.
\end{definition}

The state of a Petri net, called \emph{marking}, is a multiset of places indicating how many \emph{tokens} each place contains.
Tokens are shown as block dots inside places.

\begin{definition}[Marking]
Let $N=(P,T,F)$ be a Petri net.
A marking $M$ is a multiset of places, i.e., $M \in \bag(P)$.
\end{definition}

A transition $t \in T$ is \emph{enabled} in marking $M$ of net $N$, denoted as $(N,M)[t\rangle$, if each of its input places ($p \in{\pre t}$) contains at least one token.
An enabled transition $t$ may \emph{fire}, i.e., one token is removed from each of the input places ($p \in{\pre t}$) and
one token is produced for each of the output places ($p \in \post t$).

$(N,M)[t\rangle (N,M')$ denotes that $t$ is enabled in $M$ and firing $t$ results in marking $M'$.
Let $\sigma = \langle t_1,t_2, \ldots, t_n \rangle \in T^*$ be a sequence of transitions, sometimes referred to as a \emph{trace}.
$(N,M)[\sigma\rangle (N,M')$ denotes that there is a set of markings $M_0, M_1, \ldots, M_n$
such that $M_0 = M$, $M_n = M'$, and $(N,M_i)[t_{i+1}\rangle (N,M_{i+1})$ for $0 \leq i < n$.

A \emph{system net} has an initial and a final marking. The \emph{behavior} of a system net corresponds to the set of
traces starting in the initial marking $M_{\mi{init}}$ and ending in the final marking $M_{\mi{final}}$.

\begin{definition}[System Net Behavior]\label{def:sysnet}
A system net is a triplet $\mi{SN}=(N,M_{\mi{init}},M_{\mi{final}})$ where
$N=(P,T,F)$ is a Petri net,
$M_{\mi{init}} \in \bag(P)$ is the initial marking, and $M_{\mi{final}} \in \bag(P)$ is the final marking.
$\mi{behav}(\mi{SN}) = \{\sigma \mid (N,M_{\mi{init}})[\sigma\rangle(N,M_{\mi{final}}) \}$ is the set of traces possible according to the model.
\end{definition}

Note that a system net classifies traces $\sigma$ into \emph{fitting} ($\sigma \in \mi{behav}(\mi{SN})$) and \emph{non-fitting} ($\sigma \not\in \mi{behav}(\mi{SN})$).

\section{Hybrid Petri Nets}
\label{sec:hybridPetrinet}

A \emph{formal process model} is able to make firm statements about the inclusion or exclusion of traces, e.g.,
trace $\langle \sact,a,b,c,d,\eact \rangle$ fits the model or not. \emph{Informal process models} are unable to make such precise statements about traces.
Events logs only show example behavior:
(1) logs are typically incomplete
(e.g., the data only shows a fraction of all possible interleavings, combinations of choices, or unfoldings) and
(2) logs may contain infrequent exceptional behavior where the model should abstract from.
Therefore, it is impossible to make conclusive decisions based on event logs.
More observations may lead to a higher certainty and the desire to make a formal statement (e.g., ``after $a$ there is a choice between $b$ and $c$'').
However, fewer observations and complex dependencies create the desire to remain ``vague''.
Models (a), (b) and (c) in Figure~\ref{f-intro-prom-disco} have formal semantics as described in Definition~\ref{def:sysnet}.
(The initial and final markings are defined but not indicated explicitly: the source places are initially marked and
the sink places are the only places marked in the final markings.)
Models (d) and (e) in Figure~\ref{f-intro-prom-disco} are informal and therefore unable to classify traces into fitting and non-fitting.

In essence process models describe \emph{causalities} between activities.
Depending on the evidence in the data these causalities can be seen as stronger (``sure'') or weaker (``unsure'').
The strength of a causal relation expresses the level of confidence.
A strong causality between two activities $a$ and $b$ suggests that one is quite sure that activity $a$ causes activity $b$ to happen later in time.
This does not mean that $a$ is always followed by $b$. The occurrence of $b$ may depend on other factors, e.g., $b$ requires $c$ to happen concurrently
or $a$ only increases the likelihood of $b$.

\begin{figure}[t!]
\centerline{\includegraphics[width=9.5cm]{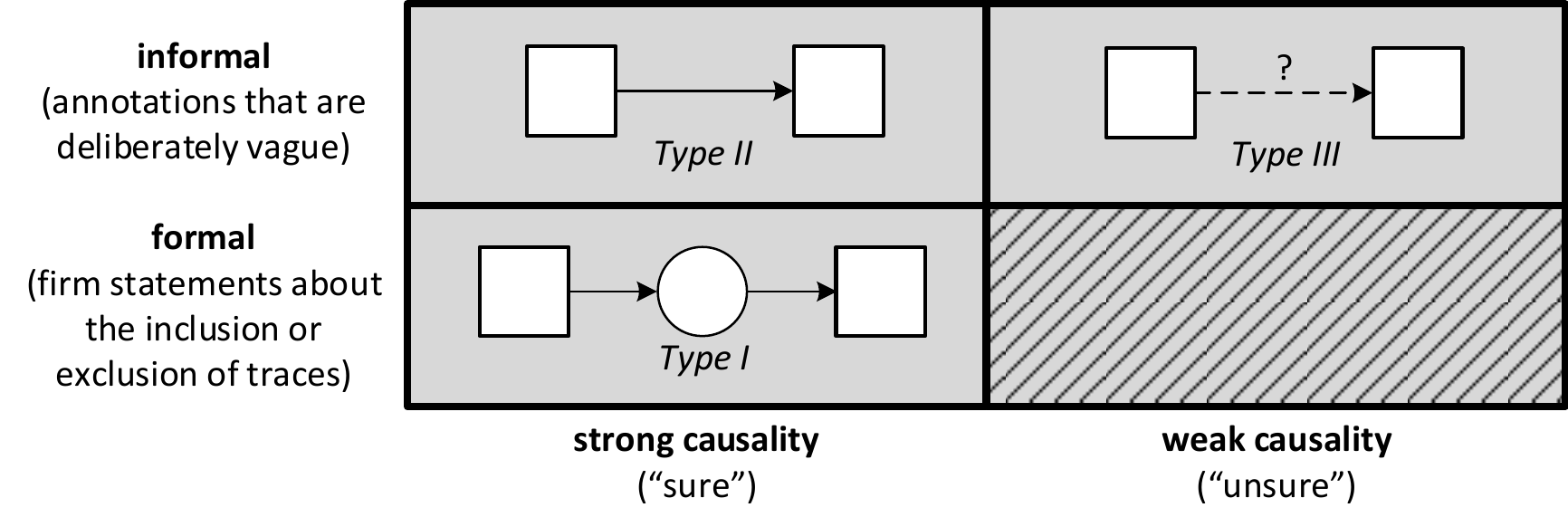}}
\caption{The strength of a causality and the formality of a modeling construct are orthogonal.
However, it makes less sense to express a weak causality in a formal manner.}
\label{fig-hybrid-pn-constructs}
\end{figure}

The strength of a causality and the formality of a modeling construct are orthogonal as shown in Figure~\ref{fig-hybrid-pn-constructs}.
Even when one is not sure, one can still use a formally specified modeling construct.
Moreover, both notions may be local, e.g., parts of the process model are more certain or modeled precisely
whereas other parts are less clear and therefore kept vague.

As Figure~\ref{fig-hybrid-pn-constructs} suggests it seems undesirable to express a weak causality using a formal construct.
Moreover, depending on the \emph{representational bias} of the modeling notation, strong causalities may not be expressed easily.
The modeling notation may not support concurrency, duplicate activities, unstructured models, long-term dependencies, OR-joins, etc.
Attempts to express behavior incompatible with representational bias of the modeling notation in a formal model are doomed to fail.
Hence, \emph{things that cannot be expressed easily in an exact manner can only be captured using annotations that are deliberately vague and non-executable.}
Instead, we aim to combine the best of both worlds, i.e.,
marrying the left-hand side and the right-hand side of Figure~\ref{f-intro-prom-disco} by combining both formal and informal notations.
\begin{figure}[t!]
\centerline{\includegraphics[width=12cm]{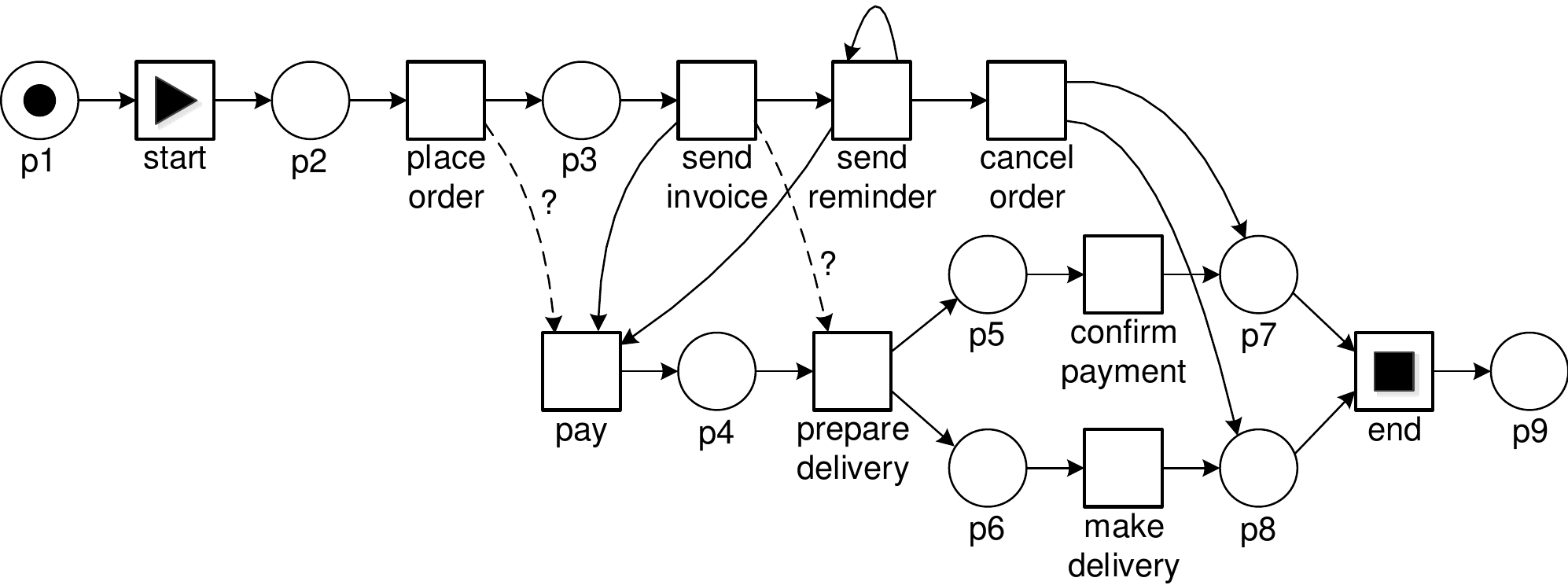}}
\caption{A hybrid system net with $M_{\mi{init}}= [p1]$ and $M_{\mi{final}}= [p9]$.
This hybrid model was discovered using the approach presented in Sect.~\ref{sec:disc}.}
\label{f-initial-hpn-example}
\end{figure}

Although the ideas are \emph{generic} and also apply to other notations (BPMN, UML activity diagrams, etc.),
we operationalize the notion of hybrid process models by defining and using so-called \emph{hybrid Petri nets}.
Unlike conventional Petri nets, we use different types of arcs to indicate the level of \emph{certainty}.

Figure~\ref{f-initial-hpn-example} shows an example of a hybrid Petri net discovered based on the event log
also used to create the models in Figure~\ref{f-intro-prom-disco}.
Strong causalities are expressed through conventional places and arcs and
\emph{sure arcs} (arcs directly connecting transitions).
Weak causalities ares expressed using \emph{unsure arcs} (dashed arcs with a question mark).
Figure~\ref{fig-hybrid-pn-constructs} shows the three types of arcs.

\begin{definition}[Hybrid Petri Net]
A \emph{hybrid Petri net} is a tuple $\mi{HPN}=(P,T,F_1,\allowbreak F_2,F_3)$ where $(P,T,F_1)$ is a Petri net, $F_2\subseteq T \times T$, and $F_3\subseteq T \times T$ such that
$\widehat{F_1}$, $F_2$, and $F_3$ are pairwise disjoint.
Arcs of Type I ($(p,t) \in F_1$ or $(t,p) \in F_1$) are the normal arcs connecting a place to a transition or vice versa.
Arcs of Type II ($(t_1,t_2) \in F_2$) are arcs indicating a strong causality between two transitions (sure arcs).
Arcs of Type III ($(t_1,t_2) \in F_3$) are arcs indicating a weak causality between two transitions (unsure arcs).
\end{definition}

Transitions, places, and normal (\emph{Type I}) arcs have formal semantics as defined in Sect.~\ref{subsec:prelim}.
Again we define an initial and final marking to reason about the set of traces possible. Therefore, we define the notion of a \emph{hybrid system net}.

\begin{definition}[Hybrid System Net]
A \emph{hybrid system net} is a triplet $\mi{HSN}=(\mi{HPN},\allowbreak M_{\mi{init}},\allowbreak M_{\mi{final}})$ where
$\mi{HPN}=(P,T,F_1,F_2,F_3)$ is a hybrid Petri net,
$M_{\mi{init}} \in \bag(P)$ is the initial marking, and $M_{\mi{final}} \in \bag(P)$ is the final marking.
${\cal U}_{\mi{HSN}}$ is the set of all possible hybrid system nets.
$\mi{behav}(\mi{HSN})$ is defined as in Definition~\ref{def:sysnet} while ignoring the sure and unsure arcs (i.e., remove $F_2$ and $F_3$).
\end{definition}

Only normal (\emph{Type I}) arcs have formal semantics; the other two types of arcs are informal and do not include or exclude traces.
Recall that Petri net without any places allows for any behavior and adding a place can only restrict behavior.
A sure arc $(t_1,t_2) \in F_2$ should be interpreted as a strong causal relationship that cannot be expressed (easily) in terms of a place connecting $t_1$ and $t_2$.
An unsure arc $(t_1,t_2) \in F_3$ is a suspected causal relationship that is too weak to justify a place connecting $t_1$ and $t_2$.

The role of sure and unsure arcs will become clearer when presenting the discovery technique in the next section.
Figure~\ref{f-initial-hpn-example} also uses special symbols for the start and end activities ($\sact$ and $\eact$) as introduced in Definition~\ref{def:eventlog},
but the semantics of $\mi{HSN}$ do not depend on this.

\section{Discovering Hybrid Process Models}
\label{sec:disc}

We aim to discover hybrid process models. As a target format we have chosen hybrid system nets that have three types of arcs.
We use a \emph{two-step approach}. First, we discover a \emph{causal graph} (Sect.~\ref{subsec:cg}).
Based on the causalities identified, we generate candidate places.
These places are subsequently evaluated using replay techniques (Sect.~\ref{subsec:fsn}).
Strong causalities that cannot be expressed in terms of places are added to the \emph{hybrid system net} as sure arcs.
Moreover, the resulting hybrid model may also express weak causal relations as unsure arcs.

\subsection{Discovering Causal Graphs}
\label{subsec:cg}

A causal graph is a directed graph with activities as nodes.
There is always a unique start activity ($\sact$) and end activity ($\eact$).
There are two kinds of causal relations: \emph{strong} and \emph{weak}.
These correspond to the two columns in Figure~\ref{fig-hybrid-pn-constructs}.

\begin{definition}[Causal Graph]
A \emph{causal graph} is a triplet $\mi{G}=(A,R_S,R_W)$ where
$A$ is the set of activities including start and end activities (i.e., $\{\sact,\eact\} \subseteq A$), $R_S \subseteq A \times A$ is the set of strong causal relations,
$R_W \subseteq A \times A$ is the set of weak causal relations,
and $R_S \cap R_W = \emptyset$ (relations are disjoint).
${\cal U}_{G}$ is the set of all causal graphs.
\end{definition}

\begin{figure}[t!]
\centerline{\includegraphics[width=8cm]{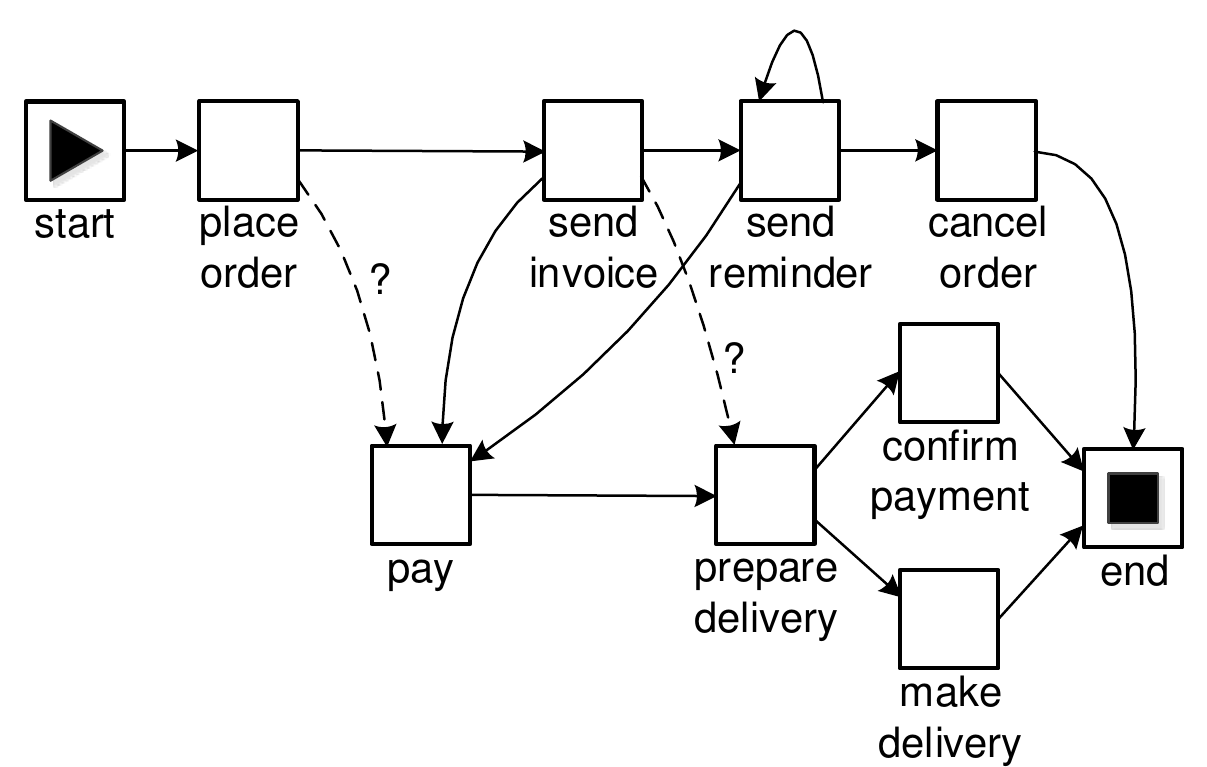}}
\caption{A causal graph: nodes correspond to activities and arcs correspond to causal relations.}
\label{f-initial-cg-example}
\end{figure}

Figure~\ref{f-initial-cg-example} shows a causal graph derived from the event log also used to discover the models in Figure~\ref{f-intro-prom-disco}.
The dashed arcs with question marks correspond to weak causal relations. The other arcs correspond to strong causal relations.

\begin{definition}[Causal Graph Discovery]
A causal graph discovery function $\mi{disc}_{\mi{cg}} \in {\cal U}_{L} \rightarrow {\cal U}_{G}$ is a function that constructs a causal graph $\mi{disc}_{\mi{cg}}(L)=(A,R_S,R_W)$ for any event log $L \in {\cal U}_{L}$ over $A$.
\end{definition}

There are many algorithms possible to construct a causal graph from a log. As an example, we use a variant of the approach used by the heuristic miner \cite{process-mining-book-2016,aal_min_icae}. We tailored the approach to hybrid discovery (i.e., different types of arcs)
while aiming for parameters that are intuitive and can be used interactively (e.g., thresholds can be changed seamlessly while instantly showing the resulting graph).
Note that we clearly separate the identification of causalities from the discovery of process logic (see Sect.~\ref{subsec:fsn}).

\begin{definition}[Log-Based Properties]\label{def:lbprop}
Let $L \in {\cal U}_{L}$ be an event log over $A$ and $\{a,b\} \subseteq A$.
\begin{itemize}[noitemsep,topsep=0pt,parsep=0pt]
  \item
$\#(a,L) = \sum_{\sigma \in L} \card{\{i \in \mi{dom}(\sigma) \mid \sigma(i) = a\}}$ counts the number of $a$'s in event log $L$.\footnote{Note that we use summations over multisets, e.g.,  $\sum_{x \in [a,b,b,a,c]} x = 2a + 2b + c$.}

\item
$\#(X,L) = \sum_{x\in X} \#(x,L)$ counts the number of $X \subseteq A$ activities in $L$.

  \item
$\#(a,b,L) = \sum_{\sigma \in L} \card{\{i \in \mi{dom}(\sigma)\setminus\{\card{\sigma}\} \mid \sigma(i) = a \ \wedge \ \sigma(i+1) = b \}}$ counts the number of times $a$ is directly followed by $b$ in event log $L$.

  \item
$\#(\ast,b,L) = \sum_{\sigma \in L} \card{\{i \in \mi{dom}(\sigma)\setminus\{\card{\sigma}\} \mid  \sigma(i+1) = b \}}$ counts the number of times $b$ is preceded by some activity.

  \item
$\#(a,\ast,L) = \sum_{\sigma \in L} \card{\{i \in \mi{dom}(\sigma)\setminus\{\card{\sigma}\} \mid  \sigma(i) = a \}}$ counts the number of times $a$ is succeeded by some activity.

  \item
$\displaystyle \mi{Rel1}(a,b,L) =  \frac{\#(a,b,L) + \#(a,b,L)}{\#(a,\ast,L) + \#(\ast,b,L)}$
counts the strength of relation $(a,b)$ relative to the split and join behavior of activities $a$ and $b$.

  \item
$\displaystyle \mi{Rel2}_c(a,b,L) =
\begin{cases}
\frac{\#(a,b,L)-\#(b,a,L)}{\#(a,b,L)+\#(b,a,L)+c} & \mbox{if} ~ \#(a,b,L)-\#(b,a,L) > 0\\
\frac{\#(a,b,L)}{\#(a,b,L)+c} & \mbox{if} ~ a=b\\
0 & \mbox{otherwise}
\end{cases}
$\\
counts the strength of relation $(a,b)$ taking into account concurrency and loops using parameter $c \in \mathbb{R}^+$ (default $c=1$).\footnote{Similar to \cite{aal_min_icae}, but negative values are mapped to 0 to get a value between 0 and 1.}

  \item
$\displaystyle \mi{Caus}_{c,w}(a,b,L) = w \cdot \mi{Rel1}(a,b,L) + (1-w) \cdot \mi{Rel2}_c(a,b,L)$
takes the weighted average of both relations where
$w \in [0,1]$ is a parameter indicating the relative importance of the first relation.
If $w=1$, we only use $\mi{Rel1}(a,b,L)$. If $w=0$, we only use $\mi{Rel2}_c(a,b,L)$.
If $w=0.5$, then both have an equal weight.

\end{itemize}
\end{definition}

$\mi{Rel1}(a,b,L)$, $\mi{Rel2}_c(a,b,L)$, and $\mi{Caus}_{c,w}(a,b,L)$ all produce values between 0 (weak) and 1 (strong).
Using the properties in Definition~\ref{def:lbprop}, we define a concrete function $\mi{disc}_{\mi{cg}}$ to create causal graphs.
All activities that occur at least $t_{\mi{freq}}$ times in the event log are included as nodes.
The strength of relations between remaining activities (based on $\mi{Caus}_{c,w}$) are used to infer causal relations.
$t_{R_S}$ and $t_{R_W}$ are thresholds for strong respectively weak causal relations.
Parameter $w$ determines the relative importance of $\mi{Rel1}$ and $\mi{Rel2}_c$.
Parameter $c$ is typically set to 1.

\begin{definition}[Concrete Causal Graph Discovery Technique]\label{def:algcg}
Let $L \in {\cal U}_{L}$ be an event log over $A$ and
let $t_{\mi{freq}} \in \Nat^+$, $c \in \mathbb{R}^+$, $w \in [0,1]$, $t_{R_S} \in [0,1]$, $t_{R_W} \in [0,1]$ be parameters such that $t_{R_S} \geq t_{R_W}$.
The corresponding causal graph is $\mi{G}=\mi{disc}_{\mi{cg}}(L)=(A',R_S,R_W)$ where
\begin{itemize}[noitemsep,topsep=0pt,parsep=0pt]
  \item $A' = \{a \in A \mid \#(a,L) \geq t_{\mi{freq}}\} \cup \{\sact,\eact\}$ is the set of activities that meet the threshold (the start and end activities are always included).
  \item $R_S = \{ (a,b) \in A' \times A' \mid \mi{Caus}_{c,w}(a,b,L\tproj_{A'}) \geq t_{R_S}\}$ is the set of strong causal relations.
  \item $R_W = \{ (a,b) \in A' \times A' \mid t_{R_S} > \mi{Caus}_{c,w}(a,b,L\tproj_{A'}) \geq t_{R_W} \}$ is the set of weak causal relations.
\end{itemize}
\end{definition}

Figure~\ref{f-initial-cg-example} shows a causal graph constructed using parameters
$t_{\mi{freq}} = 1000$, $c = 1$, $w = 0.2$, $t_{R_S} = 0.8$, and $t_{R_W} = 0.75$.

\subsection{Discovering Hybrid System Nets}
\label{subsec:fsn}

In the second step of the approach we use the causal graph to create a hybrid system net.

\begin{definition}[Hybrid System Net Discovery]
A hybrid system net discovery function $\mi{disc}_{\mi{hsn}} \in ({\cal U}_{L} \times {\cal U}_{G}) \rightarrow {\cal U}_{\mi{HSN}}$ is a function that for any event log $L$ and causal graph $G$ discovers a hybrid system net $\mi{disc}_{\mi{hsn}}(L,G) \in {\cal U}_{\mi{HSN}}$.
\end{definition}

Just like there are many algorithms possible to create a causal graph,
there are also multiple ways to construct a hybrid system net from an event log and causal graph.
The minimal consistency requirements can be defined as follows.

\begin{definition}[Consistent]\label{def:consistent}
Let $L \in {\cal U}_{L}$ be an event log, let $G=(A,R_S,R_W) \in {\cal U}_{G}$ be a causal graph, and
let $\mi{HSN}=(\mi{HPN},M_{\mi{init}},\allowbreak M_{\mi{final}}) \in {\cal U}_{\mi{HSN}}$ with
$\mi{HPN}=(P,T,F_1,F_2,F_3)$ be a hybrid system net.
$L$, $G$, and $\mi{SN}$ are \emph{consistent} if and only if:
$T = A \subseteq \bigcup_{\sigma \in L} \{a \in \sigma\}$,
$\{p_{\sact},p_{\eact}\} \subseteq P$,
$F_1 \cap ((\{p_{\sact}, p_{\eact}\} \times T) \cup (T \times \{p_{\sact}, p_{\eact}\})) = \{(p_{\sact},\sact),(\eact,p_{\eact})\}$,
$M_{\mi{init}} = [p_{\sact}]$ and $M_{\mi{final}} = [p_{\eact}]$,
for all $p \in P \setminus \{p_{\sact},p_{\eact}\}$: $\pre{p} \neq \emptyset$ and $\post{p} \neq \emptyset$,
$R_S = \widehat{F_1}\cup F_2$, $\widehat{F_1}\cap F_2 = \emptyset$, and $R_W = F_3$.
\end{definition}

An event log $L$, causal graph $G$, and hybrid system net $\mi{HSN}$
are consistent if
(1) $L$ and $G$ refer to the same set of activities all appearing in the event log,
(2) there is a source place $p_{\sact}$ marked in the initial place and enabling start activity $\sact$,
(3) there is a sink place $p_{\eact}$ marked in the final marking and connected to end activity $\eact$,
(4) all other places connect activities,
(5) there is a one-to-one correspondence between strong causal relations ($R_S$) and connections through places ($\widehat{F_1}$) or sure arcs ($F_2$), and
(6) there is a one-to-one correspondence between weak causal relations ($R_W$) and unsure arcs ($F_3$).

Consider two activities $a_1,a_2 \in A$ that are frequent enough to be included in the model. These can be related in three different ways:
$(a_1,a_2) \in \widehat{F_1}$ if there is a place connecting $a_1$ and $a_2$,
$(a_1,a_2) \in F_2$ if there is no place connecting $a_1$ and $a_2$ but there is a strong causal relation between $a_1$ and $a_2$ (represented by a sure arc),
$(a_1,a_2) \in F_3$ if there is a weak causal relation between $a_1$ and $a_2$ (represented by an unsure arc).

Any discovery function $\mi{disc}_{\mi{hsn}} \in ({\cal U}_{L} \times {\cal U}_{G}) \rightarrow {\cal U}_{\mi{HSN}}$ should ensure consistency.
In fact, Definition~\ref{def:consistent} provides hints on how to discover a hybrid system net.

Assume a place $p=(I,O)$ with input transitions $\pre p = I$ and output transitions $\post p = O$ is added.
$R_S = \widehat{F_1}\cup F_2$ implies that $\widehat{F_1} \subseteq R_S$.
Hence, $I \times O \subseteq R_S$, i.e., place $p=(I,O)$ can only connect transitions having strong causal relations.
Moreover, $I$ and $O$ should not be empty. These observations based on Definition~\ref{def:consistent} lead to the following definition of candidate places.

\begin{definition}[Candidate Places]
Let $G=(A,R_S,R_W) \in {\cal U}_{G}$ be a causal graph.
The candidate places based on $G$ are:
$\mi{candidates}(G) = \{ (I,O) \mid
I \neq \emptyset \ \wedge \
O \neq \emptyset \ \wedge \
I \times O \subseteq R_S \}$.
\end{definition}

Given a candidate place $p=(I,O)$ we can check whether it allows for a particular trace.

\begin{definition}[Replayable trace]
Let $p=(I,O)$ be a place with input set $\pre p= I$ and output set $\post p = O$.
A trace $\sigma = \langle a_1,a_2, \ldots, a_n\rangle \in A^*$ is perfectly replayable with respect to place $p$ if and only if
\begin{itemize}[noitemsep,topsep=0pt,parsep=0pt]
  \item for all $k \in \{1,2,\dots ,n\}$: $\card{\{1 \leq i < k \mid a_i \in I \}} \geq \card{\{1 \leq i \leq k \mid a_i \in O \}}$ (place $p$ cannot ``go negative'' while replaying the trace) and
 \item $\card{\{1 \leq i \leq n \mid a_i \in I \}} =$ $\card{\{1 \leq \allowbreak i \leq n \mid a_i \in O \}}$ (place $p$ is empty at end).
\end{itemize}
We write $\checkmark(p,\sigma)$ if $\sigma$ is perfectly replayable with respect to place $p=(I,O)$.
$\mi{act}(p,\sigma) = \exists_{a\in \sigma} \ a \in (I \cup O)$ denotes whether place $p=(I,O)$ has been activated, i.e., a token was consumed or produced for it in $\sigma$.
\end{definition}

Note that $\checkmark(p,\sigma)$ if $\sigma$ is a trace of the system net having only one place $p$.
To evaluate candidate places one can define different scores.

\begin{definition}[Candidate Place Scores]\label{def:scores}
Let $L \in {\cal U}_{L}$ be an event log.
For any candidate place $p=(I,O)$ with input set $\pre p = I$ and output set $\post p = O$, we define the following scores:
\begin{itemize}
  \item $\mi{score}_{\mi{freq}}(p,L) = \frac{\card{\,[\sigma \in L \mid \checkmark(p,\sigma)]\,}}{\card{L}}$ is the fraction of fitting traces,
  \item $\mi{score}_{\mi{rel}}(p,L) = \frac{\card{\,[\sigma \in L \mid \checkmark(p,\sigma) \ \wedge \ \mi{act}(p,\sigma)]\,}}{\card{\,[\sigma \in L \mid \mi{act}(p,\sigma)]\,}}$ is the fraction of fitting traces that have been activated, and
  \item $\mi{score}_{\mi{glob}}(p,L) = 1-\frac{\card{\, \#(I,L) - \#(O,L)\,}}{\mi{max}(\#(I,L),\#(O,L))}$ is a global score only looking at the aggregate frequencies of activities.
\end{itemize}
\end{definition}

To explain the three scoring functions consider again $L_1 = [
\langle \sact,a,b,c,d,\eact \rangle^{45},\allowbreak
\langle \sact,a,c,b,d,\eact \rangle^{35},\allowbreak
\langle \sact,a,e,d, \allowbreak \eact \rangle^{20}]$.
Let us consider place $p_1 = (I_1,O_1)$ with $I_1 = \{a\}$ and $O_2 = \{b\}$.
$\mi{score}_{\mi{freq}}(p_1,L_1) = \mi{score}_{\mi{rel}}(p_1,L_1)= \rfrac{80}{100} = 0.8$ and
$\mi{score}_{\mi{glob}}(p_1,\allowbreak L_1) = 1-\rfrac{\card{100 - 80}}{\mi{max}(100,80)} = 0.8$.
For place $p_2 = (I_2,O_2)$ with $I_2 = \{a\}$ and $O_2 = \{b,e\}$:
$\mi{score}_{\mi{freq}}(p_2,L_1) =  \allowbreak \mi{score}_{\mi{rel}}(p_2,L_1) = \allowbreak
\mi{score}_{\mi{glob}}(p_2,L_1) = 1$.
Hence, all three scoring functions agree and show that the second place is a better candidate.
Note that if the candidate place $p$ does not inhibit any of the traces in the log, then all scores are 1 by definition.

Let us now consider event log $L_2 = [ \langle c,d \rangle^{1000}, \langle a,b \rangle^{100}, \langle b,a \rangle^{10}, \langle a,a,a,a, \ldots\allowbreak ,a \rangle ]$ (with the last trace containing 1000 $a$'s) and candidate place $p_1 = (I_1,O_1)$ with $I_1 = \{a\}$ and $O_2 = \{b\}$.
$\mi{score}_{\mi{freq}}(p_1,L_2) = \rfrac{1100}{1111} = 0.99$,
$\mi{score}_{\mi{rel}}(p_1,L_2)~= \rfrac{100}{111} = 0.90$,
$\mi{score}_{\mi{glob}}(p_1,L_2) =
  1-\rfrac{\card{1110 - 110}}{\mi{max}(1110,110)} = 0.099$.
Now the values are very different.
Interpreting the scores reveals that $\mi{score}_{\mi{freq}}$ is too optimistic.
Basically one can add any place connected to low frequent activities, without substantially lowering the $\mi{score}_{\mi{freq}}$ score.
Hence, $\mi{score}_{\mi{rel}}$ is preferable over $\mi{score}_{\mi{freq}}$.
$\mi{score}_{\mi{glob}}$ can be computed very efficiently because traces do not need to be replayed.
It can be used to quickly prune the set of candidate places,
but the last example shows that one needs to be careful when traces are unbalanced
(i.e., $I$ or $O$ activities occur many times in a few traces).

Based on the above discussion we use scoring function $\mi{score}_{\mi{rel}}$ in conjunction with a threshold $t_{\mi{replay}}$.
The causal graph, a set of candidate places, and this threshold can be used to discover a hybrid system net.

\begin{definition}[Concrete Discovery Technique]\label{def:disctech}
Let $L \in {\cal U}_{L}$ be an event log and let $G=(A,R_S,R_W) \in {\cal U}_{G}$ be a causal graph.
$t_{\mi{replay}}$ is the threshold for the fraction of fitting traces that have been activated.
The discovered hybrid system net $\mi{disc}_{\mi{hsn}}(L,G) =(\mi{HPN},M_{\mi{init}}, M_{\mi{final}})$ with
$\mi{HPN}=(P,T,F_1,F_2,F_3)$ is constructed as follows
\begin{itemize}
  \item $Q = \{ p \in \mi{candidates}(G)  \mid \mi{score}_{\mi{rel}}(p,L\tproj_{A}) \geq t_{\mi{replay}} \}$ is the set of internal places (all candidate places meeting the threshold),
  \item $P = \{p_{\sact},p_{\eact}\} \cup Q$ is the set of places ($\{p_{\sact},p_{\eact}\} \cap Q = \emptyset$),
  \item $T = A$ is the set of transitions,
  \item $F_1 = \{(p_{\sact},\sact),(\eact,p_{\eact})\} \cup \{(t,(I,O)) \in T \times Q \mid t \in I\} \cup \{((I,O),t) \in Q \times T \mid t \in O\}$ is the set of normal arcs,
  \item $F_2 = R_S \setminus \widehat{F_1}$ is the set of sure arcs, and
  \item $F_3 = R_W$ is the set of unsure arcs.
\end{itemize}
\end{definition}

It is easy to check that this concrete $\mi{disc}_{\mi{hsn}}$ function indeed ensures consistency.
The construction of the discovered hybrid system net is guided by the causal graph.
We can construct hybrid system net $\mi{disc}_{\mi{hsn}}(L,\mi{disc}_{\mi{cg}}(L))$ for any event log $L$ using
parameters $t_{\mi{freq}}$, $c$, $w$, $t_{R_S}$, $t_{R_W}$, and $t_{\mi{replay}}$.
For example, the hybrid model shown in Figure~\ref{f-initial-hpn-example} was discovered using $t_{\mi{freq}} = 1000$, $c = 1$, $w = 0.2$, $t_{R_S} = 0.8$, $t_{R_W} = 0.75$,
and $t_{\mi{replay}} = 0.9$.
Our discovery approach is highly configurable and also provides formal guarantees (e.g., $t_{\mi{replay}}=1$ ensures perfect fitness).
When there is not enough structure or evidence in the data, the approach is not coerced to return a model that
suggests a level of confidence that is not justified.

\section{Implementation}
\label{sec:impl}

Two novel \emph{ProM} plug-ins have been created to support the approach described in this paper.\footnote{Install \emph{ProM} and the package \emph{HybridMiner} from \url{http://www.promtools.org}.}
The \emph{Causal Graph Miner} plug-in is used to create a causal graph using the approach described in Definition~\ref{def:algcg}.
The user can control the parameters $w$, $t_{\mi{freq}}$, $t_{R_S}$, and $t_{R_W}$ through sliders and directly see the effects in the resulting graph.
The \emph{Hybrid Petri Net Miner} plug-in implements Definition~\ref{def:disctech} and takes as input an event log and a causal graph.
The plug-in returns a discovered hybrid system net.
Only places that meet the $t_{\mi{replay}}$ threshold are added. The replay approach has been optimized to stop replaying a trace when
it does not fit.
\begin{figure}[t!]
\centerline{\includegraphics[width=12cm]{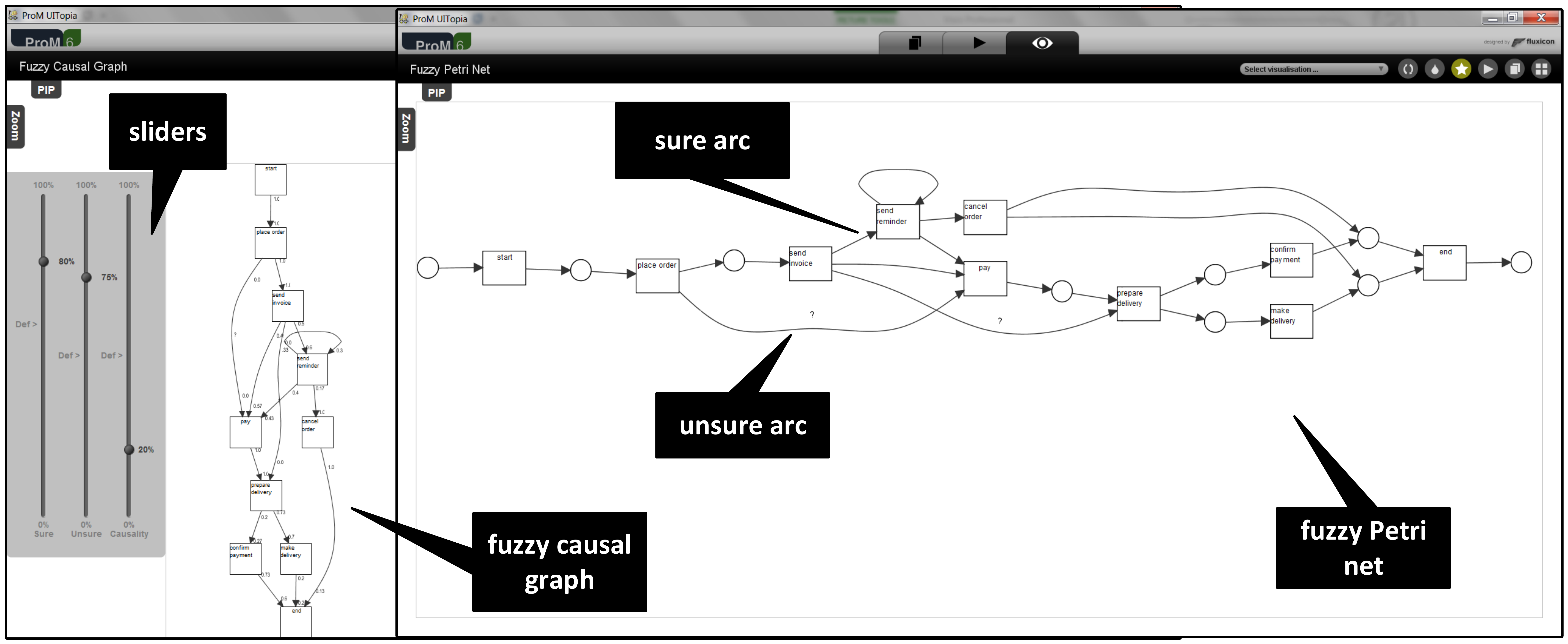}}
\caption{Screenshots of the \emph{Causal Graph Miner} (left) and the \emph{Hybrid Petri Net Miner} (right) analyzing the running example with
parameter settings $t_{\mi{freq}} = 1000$, $c = 1$, $w = 0.2$, $t_{R_S} = 0.8$, $t_{R_W} = 0.75$,
and $t_{\mi{replay}} = 0.9$.}
\label{f-tool-impl}
\end{figure}

Figure~\ref{f-tool-impl} shows the two plug-ins in action for the event log containing 12,666 cases and 80,609 events.
The results returned correspond to the causal graph depicted in Figure~\ref{f-initial-cg-example} and
the hybrid system net depicted in Figure~\ref{f-initial-hpn-example}.
\emph{Both were computed in less than a second on a standard laptop.}
Activity \emph{send reminder} may occur repeatedly (or not) after sending the invoice but before payment or cancellation.
However, payments may also occur before sending the invoice. The hybrid system net in Figure~\ref{f-tool-impl}
(also see Figure~\ref{f-initial-hpn-example} which is better readable) clearly differentiates between
(1) the behavior which is dominant and clear and (2) the more vague behavior that cannot be captured formally or is not supported by enough ``evidence''.
The example illustrates the scalability of the approach while supporting simplicity and deliberate vagueness.

\section{Evaluation}
\label{sec:eval}

In order to evaluate the quality of the discovered hybrid Petri nets we 
compute two of the classical quality dimensions used in the process discovery literature, namely \textit{replay fitness} (hereafter fitness), and \textit{precision}~\cite{genetic-mining-four-dim-coopis2012}. We computed the two metrics on six real-life datasets.

Specifically, we are interested in answering the following research questions:
\begin{itemize}
\item \textbf{RQ1}. How do fitness and precision change for different event logs?
\item \textbf{RQ2}. How do fitness and precision vary when the parameter $t_{\mi{replay}}$ vary?
\item \textbf{RQ3}. How do fitness and precision vary when the parameter $w$ changes?
\item \textbf{RQ4}. How do fitness and precision vary when the parameters $t_{R_S} $ and $t_{R_W}$ vary?
\end{itemize}

We did not perform an exhaustive comparative evaluation with classical discovery algorithms, such as the ILP miner or the Inductive Miner, as their output is an entirely formal model which is not comparable to hybrid Petri nets. Nevertheless, we carried out an overall analysis and a more qualitative comparison with state-of-the-art discovery approaches.  

In the next subsections we introduce the datasets, as well as the procedure and the metrics used for the evaluation. We finally present and discuss the obtained results. 

\subsection{Datasets, Metrics and Procedure}
The datasets used for the evaluation are six BPI Challenges. Specifically, the full BPI Challenges for 2011~\cite{BPI2011}, 2012~\cite{BPI2012} and 2017~\cite{BPI2017}, the part of the BPI Challenge 2014 concerning Activity Logs for Incidents~\cite{BPI2014-ActivityLogsIncidents}, the part of the BPI Challenge 2015 for Municipality 1~\cite{BPI2015-M1} and a subset of the BPI Challenge 2016 related to click behavior~\cite{BPI2016-complaints}. 
\begin{table}[h]
\centering
\sisetup{
table-figures-integer = 6,
table-figures-decimal = 0
}
\setlength{\tabcolsep}{6pt}
	\begin{tabular}{l r r r}
	\toprule
	\multirow{2}{*}{\textbf{Log}} & \multirow{2}{*}{\textbf{Cases}} & \multirow{2}{*}{\textbf{Events}}  & \textbf{Event} \\
	& & & \textbf{classes} \\ 	
	\midrule
	BPI 2011 & 1143  & 150291 & 624 \\ 
	BPI 2012 & 13087 & 164506 & 23  \\ 
	BPI 2014 & 46616 & 466737 & 39  \\ 
	BPI 2015 & 1199  & 52217  & 398 \\ 
	BPI 2016 & 557  & 286075 & 312  \\ 
	BPI 2017 & 31509 & 475306 & 24\\ 
	\bottomrule
	\end{tabular}
	\caption{Dataset description.}
	\label{tab:datasets}
\end{table}
Table~\ref{tab:datasets} reports the quantitative descriptions of the datasets used in the evaluation\footnote{The datasets metrics are computed after filtering the log to retain only the \textit{complete} events.}. For each dataset we report the number of cases, events, and event classes. Note that for the sake of readability we use the label BPI 2014 (resp.~2015, 2016) to denote the component of the BPI 2014 (resp.~2015, 2016) Challenge we used in the evaluation.  
 
To answer \textbf{RQ1}--\textbf{RQ4} we carried out the following steps for each of the six datasets:
\begin{enumerate}
	\item we compute a reasonable model~\footnote{We chose a \textit{reasonable} model for a given dataset, by building it so that the intermediate causal graph is connected and the resulting net has a non-trivial number of places.} for each dataset and set the values for $w$, $t_{R_S} $, $t_{R_W}$ $t_{\mi{replay}}$, and $t_{\mi{freq}}$ used as baseline values for the corresponding parameters;
	\item we compute fitness and precision based on these baseline values;  	
	\item we let $t_{\mi{replay}}$ vary using the values in $\left\{0.7, 0.8, 0.9, 1\right\}$\footnote{We chose the values of $t_{\mi{replay}}$, $w$, $t_{R_S}$ and $t_{R_W}$ based on our experience.} and we compute how fitness and precision change. 
	\item we let $w$ vary using the values in $\left\{0.0, 0.25, 0.5, 0.75, 1.0\right\}$ and we compute how fitness and precision change;
	\item we let $t_{R_S}$ and $t_{R_W}$ vary using the values in $\left\{0.5, 0.6, 0.7, 0.8, 0.9\right\}$ imposing that $t_{R_S} \geq t_{R_W}$ and we compute the corresponding trends in the number of connections through places ($\widehat{F_1}$), sure ($F_2$) and unsure ($F_3$) arcs and trends in fitness and precision;
\end{enumerate}

In order to evaluate the Hybrid Miner against the existing discovery algorithms, we filtered the five BPI dataset logs so as to leave only a limited number of event types ($\le 30$). In this way we are sure to get models with a comparable yet meaningful number of activities. We then use the logs for discovering the (hybrid) Petri nets with the different approaches and we compared the results both quantitatively (by means of fitness and precision) and qualitatively (by inspecting the discovered models).

The metrics used for computing fitness and precision are based on the state-of-the-art alignment-based approaches described in~\cite{Adriansyah2011} and \cite{Adriansyah2015}. Both metrics range between $0$ and $1$. For the fitness, the higher is its value, the more the model is able to replay the log. For the precision, the higher is its value, the fewer behaviors (i.e., traces) are possible not appearing in the event log.  In our specific setting, since a transition without any input places is always enabled, for highly ``vague'' models, i.e., models with only sure and unsure arcs, we will have (i) high values of fitness and (ii) low values of precision.

\subsection{Results}

\noindent{\textbf{7.2.1 Research Questions}}\\

Table~\ref{tab:quantitative} reports the results related to \textbf{RQ1}, that is, the replay fitness and recall of the six hybrid models discovered from the six BPI Challenges. The table reports, for each discovered hybrid model: (i) the configuration parameters provided as input (i.e., $t_{\mi{freq}}$, $t_{R_S} $, $t_{R_W}$, $w$, $t_{\mi{replay}}$)\footnote{$c$ has been set to its default value 1.}; (ii) the number of transitions, places, connections through places, sure and unsure arcs; (iii) fitness and precision of the discovered model and (iv) the time required for the computation\footnote{Times refer only to the execution of the algorithms and exclude manual input and rendering time.}. 

\begin{table}
	\centering
	\begin{scriptsize}
		\begin{tabular}{c c c c c c c c c c c c c c}
		\toprule
		 \textbf{Log} & $t_{\mi{freq}}$ & $t_{R_S} $ & $t_{R_W}$ & $w$ & $t_{\mi{replay}}$ & $|T|$ & $|P|$ & $|\mbox{$\widehat{F_1}$}|$ & $|F_{2}|$& $|F_{3}|$ & \textbf{Fitness} & \textbf{Precision} & \textbf{Time}\\ 
				\midrule
		BPI 2011 & 343  & 0.81 & 0.8  & 0.1  & 0.8  & 38 & 6  & 4  & 200 & 6  & 0.84 & 0.04   & 11772 \\ 
		BPI 2012 & 3926  & 0.9  & 0.89 & 0.1  & 0.8  & 14 & 8  & 7  & 20  & 1  & 0.9  & 0.2566 & 12414 \\
		BPI 2014 & 13985  & 0.9  & 0.9  & 0.1  & 0.8  & 10 & 5  & 3  & 13  & 0  & 0.93 & 0.535  & 21233 \\ 
		BPI 2015 & 360  & 0.45 & 0.4  &	0.5  & 0.8  & 59 & 26 & 24 & 145 & 75 & 0.74 & 0.0512 & 7055 \\ 
		BPI 2016 & 445 & 0.5  & 0.5  &	0.1 & 0.8 & 12 & 2 & 0 & 31 & 0  & 0.83 & 0.0968 & 31428 \\
		BPI 2017 & 9453  & 0.51 & 0.5  &	0.5  & 0.8  & 22 & 8  & 7  & 36  & 12 & 0.95 & 0.1227 & 24772 \\ \bottomrule
		\end{tabular}
		\end{scriptsize}
	\caption{Quantitative Evaluation}
	\label{tab:quantitative}
\end{table}

By looking at the results we can observe that the quality of the models in terms of fitness and precision changes according to the datasets analyzed. Fitness is always fairly high: its values range from a minimum of $0.74$ to a maximum of $0.96$. Precision, on the other hand, has a higher variation scope and is extremely low for the models discovered from the logs of BPI 2011, BPI 2015 and BPI 2016. These low values correspond to datasets where the discovered model is large and strongly connected. Indeed, for these datasets, the number of event classes and the trace average length is higher than for the other cases (see Table~\ref{tab:datasets}). In these cases, the $t_{\mi{replay}}$ threshold causes the discovery of a hybrid Petri net with a low number of places (with respect to the number of transitions) and a very high number of sure arcs, thus lowering down the net precision. The BPI 2016 log represents the extreme case of these scenarios. Indeed, no place has been able to exceed the $t_{\mi{replay}}$ threshold of $0.8$ (the two discovered places are the initial and the final places). This is mainly due to the high number of different event classes characterizing this log with respect to the number of cases, which makes it difficult to get evidence of strong causality relationships in the dataset. 

Instead, for datasets such as BPI 2012, and BPI 2014, the number of sure and unsure arcs is lower because the formal semantics of many causal relationships has been discovered and formalized in terms of places in the model. This results in a less ``vague'' model that allows for less extra-behaviors (with respect to the log) and hence presents a higher precision. To sum up and answer \textbf{RQ1}, we can conclude that the complexity in terms of different behaviors (and hence of relationships among the events) of a dataset has a strong impact on the quality of the discovered models, especially in terms of precision. Notice also that the algorithm computes results in up to 30 seconds.  

\begin{figure}[t!]
	\centering
		\includegraphics[width=\linewidth]{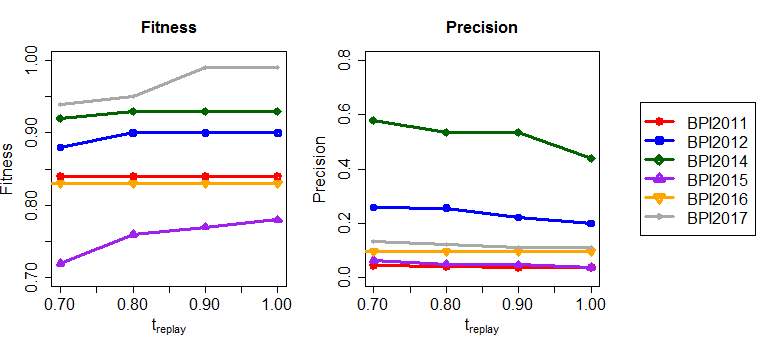}
		\caption{Fitness and precision trends for different values of $t_{\mi{replay}}$.}
		\label{fig:pEval}
\end{figure}

Figures~\ref{fig:pEval}--\ref{fig:sure} show how fitness and precision vary at different values of $t_{\mi{replay}}$, $w$, $t_{R_S}$ and $t_{R_W}$ while maintaining all the other parameters fixed according to the values reported in Table~\ref{tab:quantitative}.  

Figure~\ref{fig:pEval} shows the variation of fitness and precision for different values of $t_{\mi{replay}}$. By observing the plots, we notice that an increase in value of $t_{\mi{replay}}$ corresponds (i) for all the datasets to an increase of the fitness up to a certain threshold ($0.8$ or $0.9$) and to just a slight increase after such a threshold; (ii) a decrease of the precision value for most of the datasets (\textbf{RQ2}).
Indeed, by increasing the threshold $t_{\mi{replay}}$ we can construct, starting from the same causal graph, different hybrid Petri nets by being more and more selective on the choice of the places: only the ones ensuring a perfect fitness are discovered. This causes, on the one hand, an increase of the fitness value and, on the other hand, a decrease of the precision value, as the net contains less places. However, this is not the case for datasets as the BPI 2016 log, which have constant fitness and precision values, when varying the $t_{\mi{replay}}$ threshold. This dataset, indeed, as observed before, is characterized by a high number of different event classes with respect to the number of cases, so that only lowering the $t_{\mi{replay}}$ threshold to very low values (e.g., $0.2$) allows for finding places (different from the initial and the final place) and hence getting a variation in the values of fitness and precision.

\begin{figure}[t!]
	\centering
			\includegraphics[width=\linewidth]{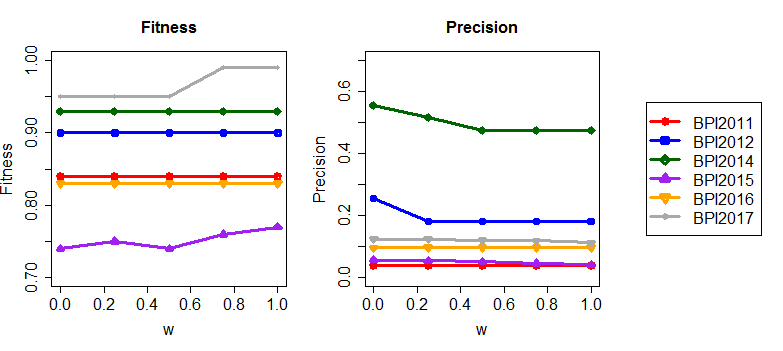}
		\caption{Fitness and precision trends for different values of $w$.}
		\label{fig:w}
\end{figure}
Focusing on $w$, the plot on the left-hand side of Figure~\ref{fig:w} shows that fitness does not change for most of the datasets: it slightly varies only for BPI 2015 and BPI 2017. This stable value of fitness can be due to the fact that the choice of the relation ($\mi{Rel1}$ or $\mi{Rel2}_{c}$) used for discovering the  causal graph does not strongly impact on the capability of the hybrid Petri net to replay the traces. In other terms, although the causality relations among activities in the causal graph can be slightly different, due to the identification of concurrency and loops, the capability of the resulting fuzzy Petri net to replay the log does not change. However, the value of $w$ has an impact on the precision: when the construction of the causal graph is done taking into account only $\mi{Rel2}_{c}$ ($w=0$), the resulting hybrid model has for some datasets (the smaller event logs with less connections among the activities) a higher precision with respect to the cases in which both relationships (or only $\mi{Rel1}$) are taken into account. This can be explained with the fact that, in the former case, concurrency and loops are identified, thus making the model more formal (or, less ``vague''). To sum up, while for fitness the impact of $w$ is minimal, for datasets discovering smaller models, the choice of considering only $\mi{Rel2}$ has a positive impact on the precision (\textbf{RQ3}).

\begin{figure}[t!]
	\centering
		\includegraphics[width=.8\linewidth]{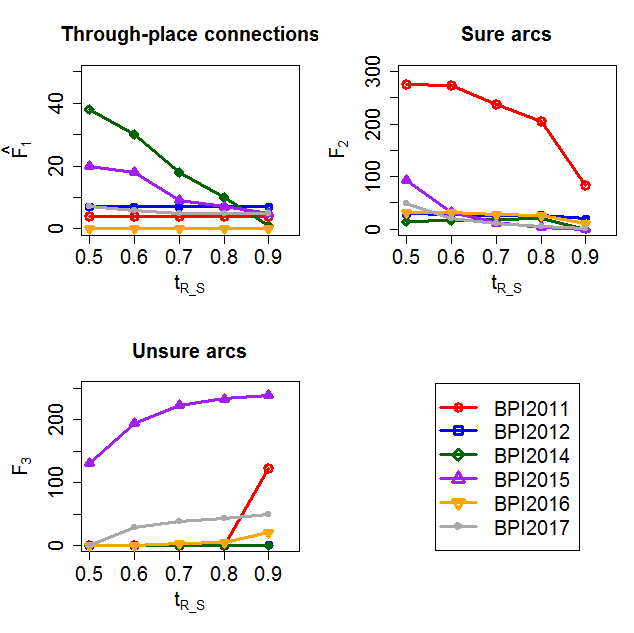}
		\caption{Connections through places, sure and unsure arc trends for different values of $t_{R_S}$ and $t_{R_W}$.}
		\label{fig:sure}
\end{figure}
The plots show how the number of connections through places, sure and unsure arcs changes for different values of $t_{R_S}$.\footnote{Note that for some of the datasets in order to guarantee $t_{R_S} \geq t_{R_W}$, we set the value of $t_{R_W}$ to the same value of $t_{R_S}$.} Overall, increasing the values of $t_{R_S}$ corresponds to a decrease in numbers of connections through places and of sure arcs, and to an increase in the number of unsure arcs. This for half of the datasets, except for the cases in which the values remain constant. This result is due to the one-to-one correspondence between the strong causal relations ($R_S$) and the connections through places ($\widehat{F_1}$) or sure arcs ($F_{2}$), so that, the higher is the $t_{R_S}$ threshold, the fewer sure arcs and connections through places occur in the model. 

\begin{figure}[t!]
	\centering
		\includegraphics[width=\linewidth]{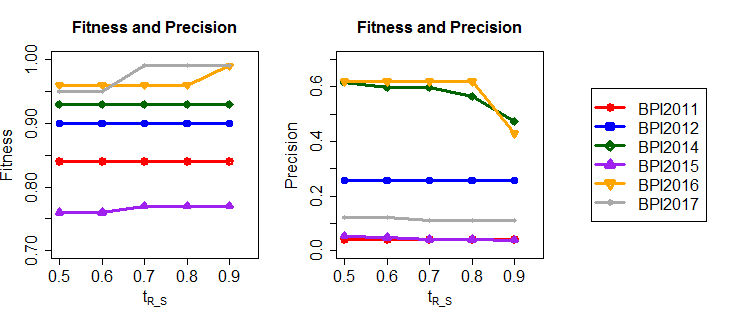}
		\caption{Fitness and precision trends for different values of $t_{R_S}$ and $t_{R_W}$.}
		\label{fig:sure}
\end{figure}

Such a trend is reflected also on fitness and precision (Figure~\ref{fig:sure}). Indeed, overall, as the number of connections through places decreases, the fitness increases or remains constant; on the contrary, the fewer places we have in the model, the more behaviors are allowed and the more the precision decreases. We omit the plots for $t_{R_W}$ as the increase of this parameter, which controls the set of unsure arcs, has no effect on fitness and precision. Similarly, an increase of this parameter has no effect on the number of connections through places and sure arcs, while we only register an expected increase of unsure arcs. (\textbf{RQ4}). 

To sum up, while the impact of $w$ is only marginal on the quality metrics, both $t_{\mi{replay}}$ (up to a value) and $t_{R_S}$ influence the quality of the discovered models, so that their increase causes a fitness increase and a precision decrease.\\

\noindent{\textbf{7.2.2 Discovery Algorithm Comparison}}\\

Moreover, we investigated the differences between the Petri nets discovered by the classical sound discovery approaches and the hybrid Petri nets returned by the Hybrid Miner. Among the discovery approaches, we focused on the Inductive Miner and the ILP miner. Unfortunately, the latter is not able to scale when applied to the investigated logs.

\begin{table}[t!]
	\centering
	\begin{scriptsize}
\begin{tabular}{l c c c c c c c c c c c c}
		\toprule
		 \textbf{Log} & \textbf{Classes} & $t_{R_S} $ & $t_{R_W}$ & $w$ & $t_{\mi{replay}}$ & $|T|$ & $|P|$ & $|\mbox{$\widehat{F_1}$}|$ & $|F_{2}|$& $|F_{3}|$ & \textbf{Fitness} & \textbf{Precision} \\ 
				\midrule
		BPI 2011 & 30 & 0.9 & 0.89 & 0.1  & 0.8  & 32 & 6  & 4  & 68 & 1  & 1 & 0.0491 \\ 
		BPI 2012 & 23 & 0.7 & 0.69 & 0.1  & 0.8  & 25 & 7  & 5  & 55  & 0  & 1  & 0.115 \\
		BPI 2014 & 30 & 0.6 & 0.59 & 0.1  & 0.8  & 32 & 8  & 10  & 170  & 1  & 0.99 & 0.2356 \\ 
		BPI 2015 & 30 & 0.5 & 0.49 & 0.4  & 0.8  & 32 & 30 & 28 & 85 & 48 & 1 & 0.1343  \\ 
		BPI 2016 & 30	& 0.4 & 0.45 & 0.1  & 0.8  & 32	& 3  & 1  & 57 & 48  & 1 & 0.2312 \\
		BPI 2017 & 24 & 0.4 & 0.39 & 0.4  & 0.8  & 26 & 9  & 8  & 115  & 5 & 0.99 & 0.1401 \\ \bottomrule
		\end{tabular}
		\end{scriptsize}
	\caption{Hybrid Miner results}
	\label{tab:hybrid}
\end{table}

\begin{table}[t!]
	\centering
	\begin{scriptsize}
\begin{tabular}{l c c c c c c c}
		\toprule
		 \textbf{Log} & \textbf{Classes} & $|T|$ & $|T_{hidden}|$ & $|P|$ & $|E|$ & \textbf{Fitness} & \textbf{Precision} \\ 
				\midrule
		BPI 2011 & 30 & 30 & 42 & 32 & 144 & 1 & 0.3172 \\ 
		BPI 2012 & 23 & 23 & 30 & 31 & 112 & 1 & 0.562 \\
		BPI 2014 & 30 & 30 & 27 & 20 & 114 & 1 & 0.452 \\ 
		BPI 2015 & 30 & 30 & 19 & 21 & 104 & 1 & 0.2584  \\ 
		BPI 2016 & 30	& 30 & 20 & 18 & 100 & 1 & 0.5735 \\
		BPI 2017 & 24 & 24 & 16 & 21 & 84  & 1 & 0.4071 \\ \bottomrule
		\end{tabular}
		\end{scriptsize}
	\caption{Inductive Miner results}
	\label{tab:inductive}
\end{table}

Table~\ref{tab:hybrid} and Table~\ref{tab:inductive} report, besides the number of event classes in the trace, the configuration settings; a description of the discovered (hybrid) Petri Net in terms of transitions, places and arcs; as well as the fitness and the precision measures for each of the two miners evaluated. Both the $t_{replay}$ threshold for the hybrid miner and the $noise$ threshold for the inductive miner have been set to 0, so as to preserve the behaviour as much as possible. Moreover, the tables also report the size of $\widehat{F_1}$, $F_{2}$ and $F_{3}$ for the hybrid miner and the number of hidden transitions, i.e., the size of $T_{hidden}$ for the inductive miner.  

\begin{figure}[h]
	\centering
			\includegraphics[width=\linewidth]{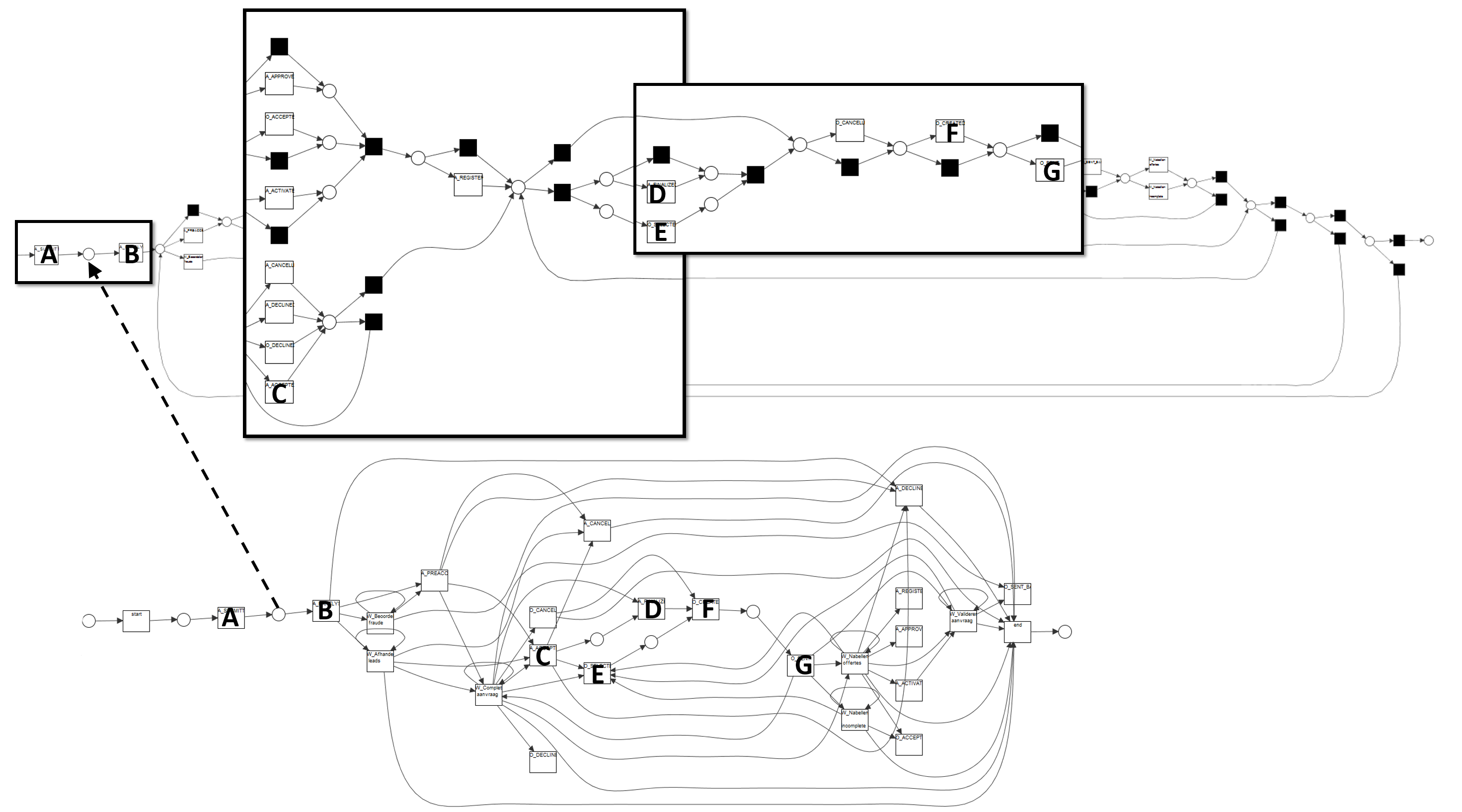}
		\caption{Mapping between the Petri nets discovered by the Inductive Miner (top) and the Hybrid Miner (bottom) from the BPI 2012 dataset}
		\label{fig:comparison2012}
\end{figure}

\begin{figure}[h]
	\centering
			\includegraphics[width=\linewidth]{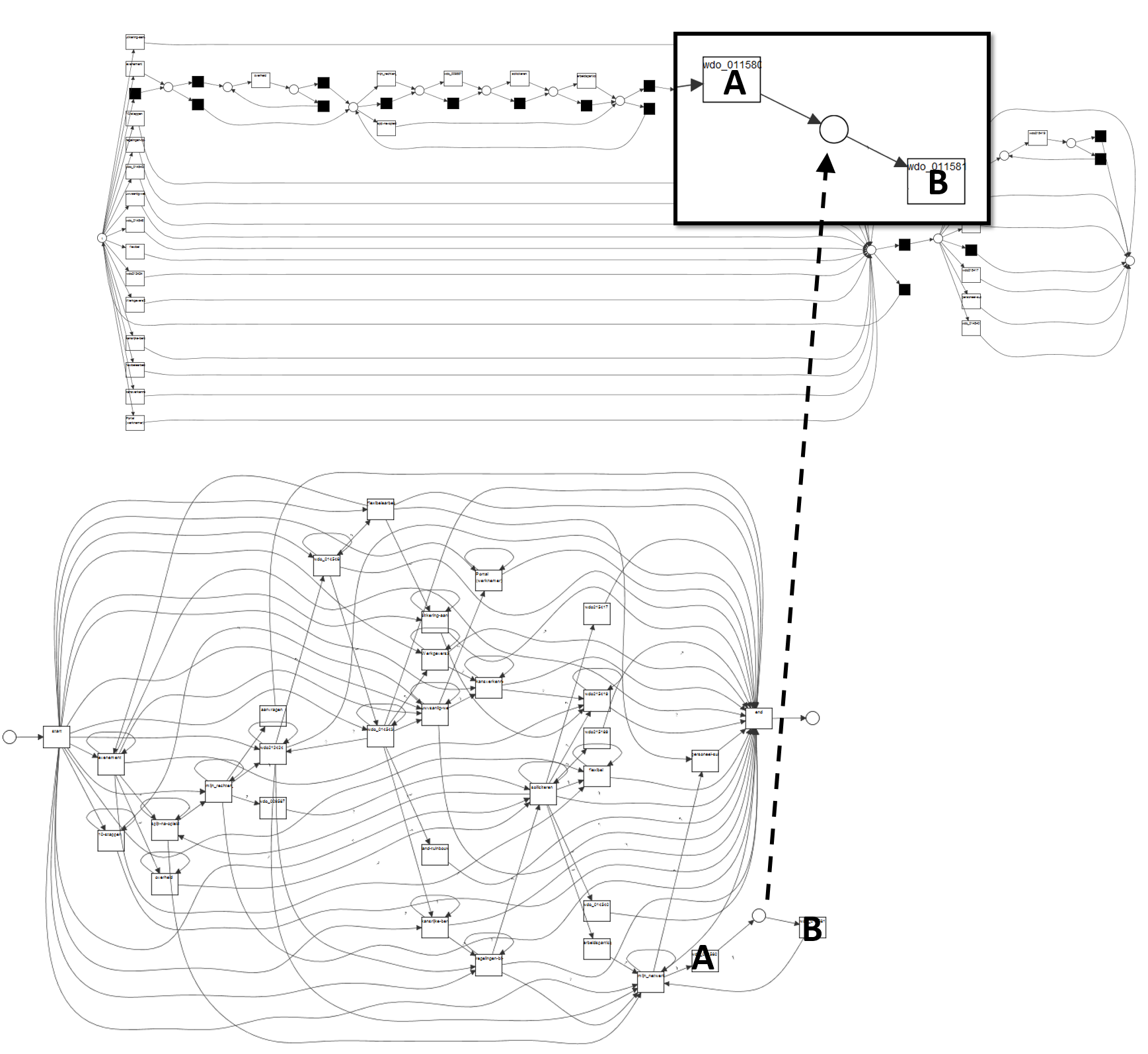}
		\caption{Mapping between the Petri nets discovered by the Inductive Miner (top) and the Hybrid Miner (bottom) from the BPI 2016 dataset}
		\label{fig:comparison2016}
\end{figure}

\begin{figure}[h]
	\centering
			\includegraphics[width=\linewidth]{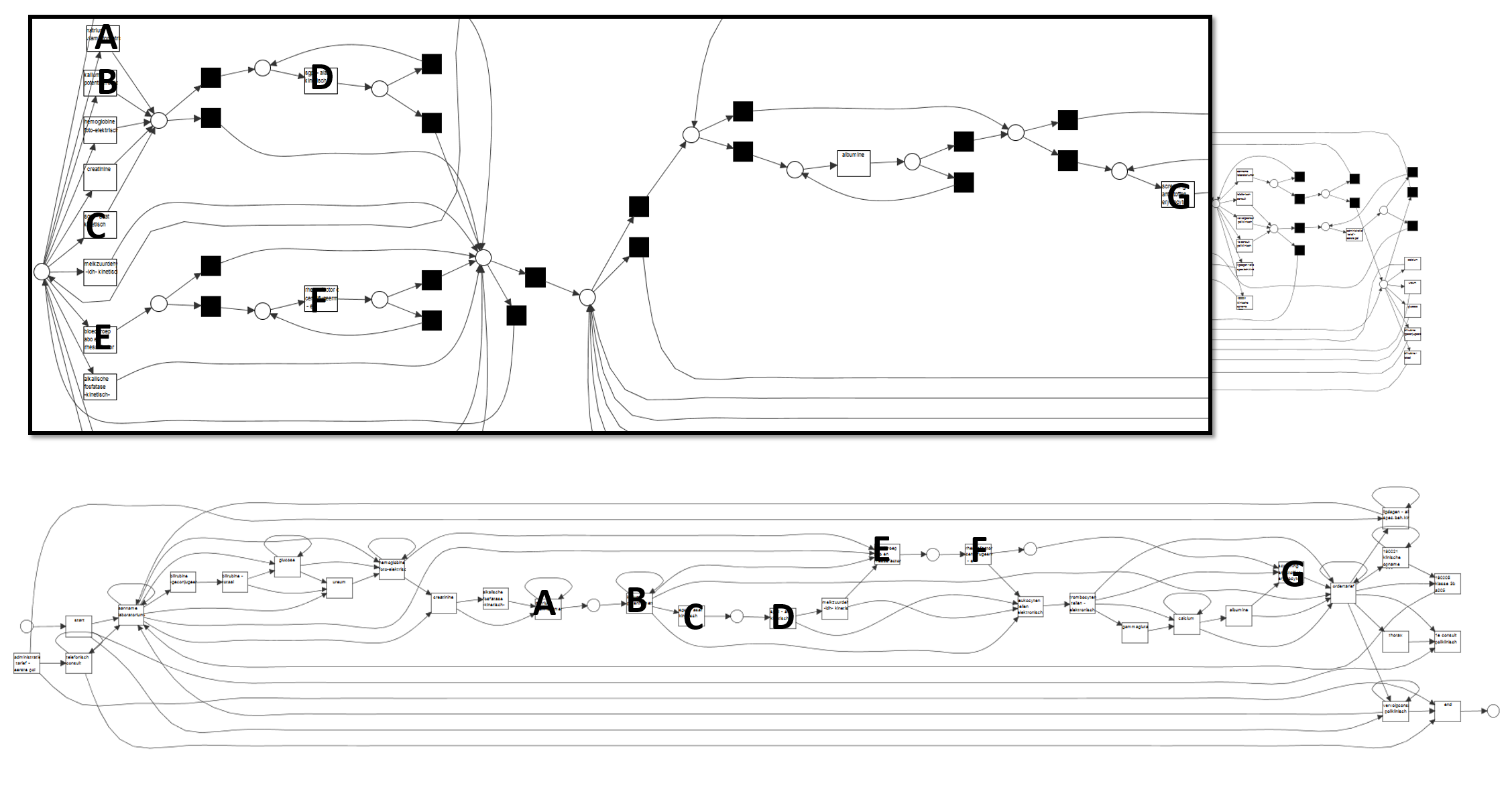}
		\caption{Mapping between the Petri nets discovered by the Inductive Miner (top) and the Hybrid Miner (bottom) from the BPI 2011 dataset}
		\label{fig:comparison2011}
\end{figure}

By comparing the results of the two tables, we can observe that the Inductive Miner is able to provide results with a similar fitness and with an improved precision. However, 
the Petri net returned by the Inductive miner, contains several hidden transitions, which could bring to log underfitting.

A qualitative inspection of the two nets, indeed, reveals that while some of the places discovered by the Hybrid miner (and hence intuitively indicating that in $80\%$ of the cases the place is able to perfectly replay the log, as $t_{replay}$ is set to $0.8$) have a corresponding place in the net discovered by the Inductive Miner, in other cases the relationship identified by the hybrid Petri net is not captured by the Petri net discovered by the Inductive Miner. Indeed, many of the places in the Petri net discovered by the Inductive miner are not places with a sequential, parallel or alternative semantics, because of the hidden transitions. This is clearly shown in Figure~\ref{fig:comparison2012}, which reports one of the nets discovered for the BPI 2012 dataset by the Inductive and the Hybrid Miner, respectively. In this case, only one of the relationships identified by the Hybrid Miner (the one between the activities $A$ and $B$) is also captured by the Inductive miner. All the others (e.g., the pair $C$ and $D$ and the pair $E$ and $F$) are actually not identified by the Inductive Miner because of the hidden transitions. This is even more evident in the results related to the BPI 2016. Figure~\ref{fig:comparison2016} shows the nets discovered from the logs by the two algorithms. Despite the high number of places discovered by the Inductive Miner ($18$ as reported in Table~\ref{tab:inductive}), only one of them expresses a clear causality relation, i.e., whenever the activity $A$ is observed, the activity $B$ has to be observed as well. All other internal places are part of the so-called ``flower loop''. The only meaningful place is precisely the place that is also discovered by the Hybrid Miner. Similarly, in the Petri net discovered by the Inductive Miner from the BPI 2011 dataset (reported in Figure~\ref{fig:comparison2011}), none of the $32$ places expresses a clear causality relations because of the hidden transitions. On the contrary, the Hybrid Miner is able to identify, apart from the initial and the final places, $4$ meaningful places (connecting the pair $A$-$B$, the pair $B$-$C$, the pair $D$-$E$ and the pair $E$- $F$). 
These results suggest hence that existing precision metrics fail to  compute precision well in the presence of many silent transitions.

\section{Related Work}
\label{sec:relwork}

The work reported in this paper was inspired by the work of Herrmann et al.\  \cite{herrmann2000,herrmann-bit-1999} who argue that modeling
``requires the representation of those parts of knowledge which
cannot be stated definitely and have to be modeled vaguely''.
They propose annotations to make vagueness explicit.
In \cite{herrmann2000,herrmann-bit-1999} the goal is to \emph{model} vagueness,
but we aim to automatically \emph{discover} hybrid models
supporting both vagueness and formal semantics.

Hybrid process models are related to the partial models considered in software engineering \cite{Famelis-ICSE2012,Salay-RE2013}.
However, these partial models are closer to configurable process models representing sets of concrete models.

In literature one can find a range of process discovery approaches
that produce formal models \cite{process-mining-book-2016}.
The $\alpha$-algorithm \cite{aal_min_TKDE} and its variants produce a Petri net.
Approaches based on state-based regions \cite{carmona-PN2010}
and language-based regions \cite{lorenz_BPM2007,language_mining_dongen-fundamenta2009}
also discover Petri nets. The more recently developed inductive mining approaches
produce process trees that can be easily converted to Petri nets or similar
\cite{sander-tree-disc-PN2013,sander-infreq-bpi2013-lnbip2014,sander-scalable-procmin-SOSYM}.

Commercial process mining tools typically produce informal models.
These are often based on the first phases of the heuristic miner \cite{aal_min_icae} (dependency graph)
or the fuzzy miner \cite{fuzzy_BPM2007} (not allowing for any form of formal reasoning).

It is impossible to give a complete overview of all discovery approaches here.
However,as far as we know there exist on other discovery approaches that return hybrid models having both formal and informal elements.

\section{Conclusion}
\label{sec:concl}

In this paper we advocated the use of \emph{hybrid models} to combine the best of two worlds: commercial tools producing informal models
and discovery approaches providing formal guarantees.
We provided a concrete realization of our hybrid discovery approach using \emph{hybrid Petri nets}.
The ideas are not limited to Petri nets and could be applied to other types of
process models (e.g., BPMN models with explicit gateways for the clear and dominant behavior and
additional arcs to capture complex or less dominant behavior).
Unlike existing approaches there is no need to straightjacket behavior into a formal model
that suggests a level of confidence that is not justified.
The explicit representation of vagueness and uncertainty in hybrid process models is analogous to
the use of confidence intervals and box-and-whisker diagrams in descriptive statistics.

The approach has been fully implemented and tested on numerous real-life event logs.
The results are very promising, but there are still many open questions. In fact,
the paper should be seen as the starting point for a new branch of research in
BPM and process mining. Future work will include instantiations of the approach
for BPMN and UML activity diagrams focusing on different model constructs (gateways, swimlanes, artifacts, etc.).
Existing techniques (also supported by \emph{ProM}) can already be used to map compliance and performance indicators
onto causalities expressed in terms of explicit places.
We would like to also provide approximative compliance and performance indicators for sure and unsure arcs.
Note that commercial tools show delays and frequencies on arcs, but these indicators may be very misleading as
demonstrated in Sect.~11.4.2 of \cite{process-mining-book-2016}.
Finally, we would like to improve performance.
The approach has already a good performance. Moreover, there are several ways to further
speed-up analysis (e.g., pruning using $\mi{score}_{\mi{glob}}$ or user-defined preferences).
Moreover, computation can be distributed in a straightforward manner (e.g., using MapReduce).

\bibliographystyle{plain}
\bibliography{lit}

\end{document}